\documentclass{elsart}
\usepackage{latexsym}
\usepackage{longtable}
\usepackage{dcolumn}
\newcolumntype{.}{D{.}{.}{-1}}

\begin{document}
\begin{frontmatter}
\journal{J. Mol. Struct. (Theochem)}
\title{Correlation Energy Estimators based on M{\o}ller-Plesset
Perturbation Theory}
\author{Herbert H. H. Homeier\thanksref{HHHH}}
\address{
 Institut f\"ur Physikalische und Theoretische Chemie,
 Universit\"at Regensburg,
 D-93040 Regensburg, Germany}
\thanks[HHHH]{na.hhomeier@na-net.ornl.gov, %
http://rchs1.uni-regensburg.de/\%7Ec5008/}

\begin{abstract}
Some methods for the convergence acceleration of the M{\o}ller-Plesset
perturbation series for the correlation energy are discussed. The
order-by-order summation is less effective than the Feenberg series. The
latter is obtained by renormalizing the unperturbed Hamilton operator
by a constant factor that is optimized for the third order energy. In
the fifth order case, the Feenberg series can be improved by
order-dependent optimization of the parameter. Alternatively, one may
use Pad\'e approximants or a further method based on effective
characteristic polynomials to accelerate the convergence of the
perturbation series. Numerical evidence is presented that, besides the
Feenberg-type approaches, suitable Pad\'e approximants, and also the
effective second order characteristic polynomial, are excellent tools
for correlation energy estimation. %
\end{abstract}
\end{frontmatter}
                     \section{Introduction}

Quite often in theoretical work, approximation schemes for some
quantities converge rather slowly.   Thus, there is a need for means to
accelerate convergence or, equivalently, to extrapolate from few
members of a sequence to its limit. Fortunately, the development of
such methods has become a rather active field at the borderline between
mathematics and the sciences in recent years. Brezinski and Redivo
Zaglia \cite{BrezinskiRedivoZaglia91} have given an excellent
mathematical introduction to such methods. There are many methods that
can be used to accelerate slowly convergent (or to sum divergent) power
series in terms of rational approximations, e.g., Pad\'e approximants
\cite{Baker65,BakerGravesMorris81a,BakerGravesMorris81b} that are
related to the famous epsilon algorithm \cite{Wynn56a}, Levin-type
methods \cite{Levin73,Weniger89,Homeier93,HomeierWeniger95}, and
iterative methods \cite{Weniger91} like the recently developed
$\mathcal{J}$ transformation
\cite{Homeier94a,Homeier94d,Homeier94e}. There are also methods
that can be used to accelerate the convergence of Fourier
\cite{KieferWeiss81,Homeier92,Homeier93} and other orthogonal
series \cite{Longman87,Homeier94b}. Onedimensional iteration
sequences can be accelerated very effectively as is demonstrated
in \cite{Homeier95} for the case of the inverse Dyson equation.
There is also a growing literature on extrapolation of matrix
and vector sequences (see \cite{BrezinskiRedivoZaglia91} for an
introduction) that have found applications to the computation of
matrix functions \cite{Homeier94c} and the iterative
solution of fixed-point equations \cite{HomeierRastKrienke95}.
The full potential for application of these methods in the
sciences has still to be explored.

One of the fields where these methods may be applied is
Many-Body Perturbation Theory (MBPT), that is  one of the standard methods to
obtain the correlation energy in molecular {\em ab initio}
calculations. The convergence acceleration of many-body perturbation
series has recently become a topic of increasing interest
\cite{SchmidtWarkenHandy93,DietzSchmidtWarkenHess92,%
DietzSchmidtWarkenHess93a,DietzSchmidtWarkenHess93b,%
DietzSchmidtWarkenHess93c,DietzSchmidtWarkenHess94a,%
DietzSchmidtWarkenHess94b}, also in the context of time-dependent
phenomena \cite{DietzSchmidtWarken95}. Here, we restrict attention to
approaches to correlation energy estimation that are based on the
M{\o}ller-Plesset (MP) series since the latter is commonly and
routinely used in quantum chemistry for closed-shell systems. For
open-shell systems, the restricted MP (RMP) method has been developed
\cite{KnowlesAndrewsAmosHandyPople91} that is based on an restricted
open-shell Hartree-Fock (ROHF) determination of the MP unperturbed
Hamiltonian. In this way, the RMP approach largely avoids spin
contaminations that are characteristic for unrestricted MP (UMP) based
on an unrestricted HF (UHF) zero-order calculation. For smaller
molecules, calculations up to fourth or even fifth order do not pose
large problems, and MPn (n=2,4) calculations are a popular approach to
the correlation problem. However, the computational effort increases
steeply with the order of the perturbation series, and with the size of
the molecular system. Therefore, there is a need to make the best use
of the lower-order terms since higher terms are difficult to obtain.
Order-by-order summation of the perturbation expansion as given by
\begin{equation}\label{eq0}
E = E_{0} + E_{1} + E_{2} + E_{3} +  E_{4}
+  E_{5} + \dots\>,
\end{equation}
i.e., using the $n$th order estimate
\begin{equation}\label{eq0a}
E^{(n)} = \sum_{j=0}^{n} E_j\>,
\end{equation}
is not the best way to exploit the information content
of its terms. It has been shown by Schmidt, Warken and Handy
\cite{SchmidtWarkenHandy93} that a specific variant of a method
originally proposed by Goldhammer and Feenberg
\cite{GoldhammerFeenberg56,Feenberg56} for the Brillouin-Wigner
perturbation expansion allows to obtain better estimates for the
correlation energy than order-by-order summation of the usual MP series.
This variant was called the {\em Feenberg series} in
\cite{SchmidtWarkenHandy93}. It is also a special case of the so-called
{\em Geometric Approximation}
\cite{Amos70,Bhattacharyya81,SchulmanMusher68,WilsonSilverFarrell77}.
Similar to the original approach of Goldhammer and Feenberg
\cite{GoldhammerFeenberg56,Feenberg56}, the computation of the Feenberg
series requires only the terms $E_{j}$ of the perturbation series.

Alternatively, one may use Pad\'e approximants
that provide rational approximations $[p,q]$ to power series,
where $p$ denotes the order of the numerator polynomial, and $q$
that of the denominator polynomial.  Pad\'e approximants may be
calculated for the original perturbation series, and also for
the renormalized perturbation series. As shown by Wilson,
Silver, and Farrell \cite{WilsonSilverFarrell77}, the special
Pad\'e approximants $[n+1,n]$ have the property that they are
invariant under the scaling of the unperturbed Hamilton operator
and, thus, are identical for the original and the renormalized
case. This invariance is an important property of correlation
energy estimators since the true correlation energy is
independent of our choice of the unperturbed Hamiltonian.

Recently, a method based on effective characteristic polynomials
has been applied to correlation energy computations of some model
systems
\cite{Bracken94,BrackenCizek94,TakahashiBrackenCizekPaldus95,%
BrackenCizek95a,BrackenCizek95b,CizekBracken95,%
DowningMichlCizekPaldus79}
and for the summation of perturbation expansions of anharmonic
oscillators \cite{CizekWenigerBrackenSpirko96}. We will see that
results based on low-order effective characteristic polynomials
also have the desirable invariance property under rescaling of
the unperturbed Hamiltonian.

All these methods require only the terms $E_{i}$ of the
M{\o}ller-Plesset perturbation series. The additional effort to
calculate them besides the usual perturbation series is very
low. As will be shown, these methods allow to obtain much better
estimates of the correlation energy in many cases, and allow the
identification of cases where standard perturbation theory
fails. In these cases, computationally more demanding
correlation energy estimators have to be used \cite{%
SchmidtWarkenHandy93,DietzSchmidtWarkenHess92,%
DietzSchmidtWarkenHess93a,DietzSchmidtWarkenHess93b,%
DietzSchmidtWarkenHess93c,DietzSchmidtWarkenHess94a,%
DietzSchmidtWarkenHess94b,%
Cizek69,Paldus74,%
Paldus76,%
Kutzelnigg77,%
Shavitt77,%
HoseKaldor79,%
Bartlett81,%
JeziorskiMonkhorst81,%
HoseKaldor82,%
BartlettDykstraPaldus84,%
Wilson84,%
HoffmanSchaefer86,%
Hose89,%
Paldus88,%
MukherjeePal89,%
SzaboOstlund89,%
Fulde91,%
Karwowski92,%
McWeeny92,%
Paldus92,%
Roos92,Roos94,%
Handy94,%
Rychlewski94,%
Meissner95,%
SteinbornSilver96}.

                       \section{Methods}
The Goldhammer-Feenberg approach \cite{GoldhammerFeenberg56,Feenberg56}
renormalizes the unperturbed Hamiltonian $H_0$ by a constant factor
according to
\begin{equation}\label{eq1}
H_0(\alpha) = (1-\alpha) H_0\>.
\end{equation}
This leads to a repartitioning of the total Hamiltonian
$H=H_0+H_1$ as
\begin{equation}\label{eq0alpha}
H = H_0(\alpha) + H_1(\alpha), \quad H_1(\alpha)= H_1 + \alpha H_0\>.
\end{equation}
It also leads to a renormalized perturbation series
\begin{equation}\label{eq3}
E(\alpha) = E_{0}(\alpha) + E_{1}(\alpha) + E_{2}(\alpha) +
E_{3}(\alpha) +  E_{4}(\alpha) +  E_{5}(\alpha) +
\dots\>
\end{equation}
with partial sums --- i.e., renormalized $n$th order energy
estimates --- given by
\begin{equation}\label{eq3a}
E^{(n)}(\alpha) = \sum_{j=0}^{n} E_j(\alpha)\>
\end{equation}
depending on renormalized $j$th order contributions \cite[Eq.
(12)]{Feenberg56}
\begin{equation}\label{eq5}
\begin{array}{rl}
E_{0}(\alpha) =&\displaystyle (1-\alpha) E_{0}\>, \qquad
E_{1}(\alpha) =\displaystyle E_{1} + \alpha E_{0}\>,\\
E_{n}(\alpha) =&\displaystyle
\frac{1}{(1-\alpha)^{n-1}} \,\sum_{j=2}^{n}
{{n-2}\choose{j-2}}(-\alpha)^{n-j} \, E_{j}\>,\qquad (n\ge 2)\>.
\end{array}
\end{equation}
For the Feenberg series, the factor $\alpha$ is determined by
requiring that the third order energy $E^{(3)}(\alpha)$
of the renormalized perturbation expansion is stationary with respect
to variations of the factor $\alpha$. This leads to an optimized
value based on the third order result given by
$\alpha^{(3)}=E_{3}/E_{2}$.
In this way, the partitioning of the Hamiltonian is fixed, and the
Feenberg series is obtained as the usual Rayleigh-Schr\"odinger series
for the unperturbed Hamilton operator $H_0(\alpha^{(3)})$.
The total energies are
\begin{equation}\label{eqFn}
F_n = E^{(n)}(\alpha^{(3)}) = E^{(n)}(E_{3}/E_{2})\>.
\end{equation}

The stationarity of the eigenvalue is based on the observation
that the exact value of the energy, i.e., the infinite order
result should be independent of the value of $\alpha $ that is
used. When applying this to an approximation obtained in some
finite order, that value of $\alpha$ is best where the
derivative of the approximation is as small as possible in
absolute value, preferably zero. We remark that this is related
to the concept of order-dependent mappings as discussed in
\cite[Sec. 18]{ArtecaFernandezCastro90}. Since order-by-order
summation of the $\alpha$ dependent Rayleigh-Schr\"odinger
expansion leads to the $n$th order estimate $E^{(n)}(\alpha)$
defined in Eq.~(\ref{eq3a}), the optimal value $\alpha^{(n)}$ of
$\alpha$ in $n$th order is determined from the equation $(n>
1)$
\begin{equation}\label{eq8}
0=\frac{d E^{(n)}} {d \alpha} (\alpha^{(n)})\>, \qquad \frac{d
E^{(n)}} {d \alpha} (\alpha)
 = (n-1)\, E_n(\alpha) / (1-\alpha) \>.
\end{equation}
The second equality here follows from an explicit calculation.
A solution of this equation leads to an approximation
\begin{equation}\label{eqGFn}
GF_n=E^{(n)}(\alpha^{(n)})
\end{equation}
for the total energy. Thus, in each order of the
renormalized perturbation series, different values of $\alpha$
are
chosen. This approach has been proposed already by Feenberg. We will
call its results the total {\em Goldhammer-Feenberg
energies}
in order to distinguish it from the Feenberg total energies. Obviously,
there can be several solutions of Eq. (\ref{eq8}), and the
Goldhammer-Feenberg energies are not guaranteed to be real.

In the case
of fifth order, the condition (\ref{eq8}) reduces in combination
with Eq.~(\ref{eq5}) to requiring that
$\alpha^{(5)}$ is a root of the third order polynomial
$(1-\alpha)^4\,E_5(\alpha)$. The latter has
real coefficients and, thus, is guaranteed to have a real
solution $\alpha^{(5)}_r$. The corresponding value $E^{(5)}(\alpha^{(5)}_r)$ will be called
GF5 later. Alternatively, one can use the average of the two (in
the present case always) complex energies obtained from the
other roots of the third order polynomial. This average will be
called GF5b later.

As is well-known (see for instance \cite{Kutzelnigg77,Hose89}),
Rayleigh-Schr\"odinger MBPT is size-extensive order by order,
i.e., for a super-molecule build up from $N$ non-interacting
identical systems, the perturbation energies are linear in $N$
in each order. Thus, if $E_j$ is the $j$th term of the
perturbation series of one of the $N$ subsystems, the $j$th
order term of the perturbation series for the super-molecule is
$N\,E_j$.

In the case of the Feenberg scaling, we note that Eq.\
(\ref{eq5}) implies that for $E_j\to N\,E_j$, we also have
$E_j(\alpha)\to N\,E_j(\alpha)$. Thus, for any $\alpha$ that
is independent of $N$, also
the renormalized perturbation series is size-extensive in each
order. Since
$\alpha^{(3)}=E_3/E_2$ is invariant under $E_j\to N\,E_j$, all Feenberg
energies $F_n$ are size-extensive as a consequence of Eq.\
(\ref{eqFn}).

The Goldhammer-Feenberg energies
$GF_n$ for $n>1$ are also size-extensive. To prove this, we note that
under $E_n\to N\,E_n$, we have $d E_n/d \alpha\to N\,d
E_n/d\alpha$. This follows from the last equality in Eq.\ (\ref{eq8}), since
$E_n(\alpha)\to N\,E_n(\alpha)$ under $E_n\to N\,E_n$.
This implies that the positions of the zeros of $d E_n/d \alpha$,
and hence the positions $\alpha^{(n)}$ of the extrema of
$E_n(\alpha)$ are invariant under $E_n\to N\,E_n$. Since the
$\alpha^{(n)}$ are used to define the Goldhammer-Feenberg energies, the
latter  are size-extensive. In particular, this applies to GF5
and GF5b.

  Now, we sketch the method of the effective characteristic polynomial
  that has recently been applied to the summation of divergent
  perturbation series \cite{CizekWenigerBrackenSpirko96}. In the linear
  variation method with $n$ orthonormal basis functions
  $\{\phi_j\}_{j=1}^{n}$ applied to a Hamiltonian $H$, the
  characteristic polynomial $P_n(E)$ of degree $n$ in the unknown energy
  $E$  has the form
\begin{equation}\label{eqC1}
P_n(E)={\rm det}\,\left\vert \left\langle \phi_j \vert H \vert
\phi_k \right\rangle -E\,\delta_{j,k}\right\vert \>.
\end{equation}
  If $H=H_0+\beta V$, the polynomial has the form
  (\cite{CizekWenigerBrackenSpirko96}, Eq. (3.2))
\begin{equation}\label{eqC2}
P_n(E) = \sum_{j=0}^{n} E^j \,\sum_{k=0}^{n-j} f_{n,j,k} \beta^k
\end{equation}
  with $f_{n,n,0}=1$. Thus, $N=n(n+3)/2$ coefficients $f_{n,j,k}$ have
  to be determined. They could be obtained
  from the matrix elements of $H_0$ and $V$. In the method of the
  characteristic polynomial, they are obtained from the coefficients of
  the perturbation series for $E$
\begin{equation}\label{eqC3}
E = \sum_{j=0}^{\infty} E_j \, \beta^j\>.
\end{equation}
  For this end, one uses (\ref{eqC3}) in (\ref{eqC2}) and does a Taylor
  expansion in $\beta$ with the result
\begin{equation}\label{eqC4}
P_n\left(\sum_{j=0}^{\infty} E_j \, \beta^j\right) = \sum_{k=0}^{N-1}
A_k \beta^k + O\left(\beta^{N}\right)\>.
\end{equation}
The $A_k$ depend on the $f_{n,j,k}$. Since $P_n(E)=0$ for an eigenvalue
$E$, one demands
\begin{equation}\label{eqC5}
A_k = 0\>, \qquad  0\le k \le N-1\>.
\end{equation}
This yields a linear equation system for the unknown
$f_{n,j,k}$, and thus, these coefficients can be determined.
After the determination, the effective characteristic equation
$P_n(E)=0$ is
solved for $E$. If only perturbation coefficients $E_j$ up to $j=5$ are
available, only a second degree effective characteristic
polynomial can be used. In our case, one finally puts $\beta=1$. In this way,
one obtains an explicit solution of $P_2(E)=0$ as
\begin{equation}\label{eqP2}
 \Pi_2 = E_0 + E_1  +
 {\displaystyle \frac {E_2^2}{2}}
 \,\frac{\displaystyle
    E_2 -E_3 +
   \sqrt {(E_2- E_3)^2 - 4 \,(E_2\,E_4- E_3^2)}
   }
   {\displaystyle  E_2\,E_4 - E_3^2
   }
\end{equation}
A further solution (with a minus sign of the square root) only
yields the correct result for small $\beta$ if $E_2>0$
holds which does not occur in perturbation theory calculations of
ground states.

Direct calculation shows that the estimate $\Pi_2$ is
independent under a scaling of $H_0$, i.e., we have
\begin{equation}\label{eqP2a}
\Pi_2(E_0,\dots,E_4) = \Pi_2(E_0(\alpha),\dots,E_4(\alpha))\>.
\end{equation}
Since the true characteristic polynomials --- depending only on
the total Hamiltonian --- are invariant under Feenberg scaling,
it may be conjectured that this invariance also holds for
estimates obtained as roots of effective characteristic
polynomials of higher degree. A proof of this conjecture is
under investigation.

We denote ${\Pi}_{2}$  also as estimate $\Pi$2 for the total
energy in the following.

It is easy to see from Eq.\ (\ref{eqP2}) that $\Pi_2\to
N\,\Pi_2$ if $E_j\to N\,E_j$ for all $j$ with $0\le j\le 4$. Thus, the
$\Pi_2$ estimator is size-extensive.

Pad\'e approximants
\cite{Baker65,BakerGravesMorris81a,BakerGravesMorris81b} are defined
with respect to a given power series as ratios of two polynomials.
Given numerator and denominator polynomial degrees $p$ and $q$, the
coefficients of these polynomials in the Pad\'e approximant $[p,q]$ are
determined by requiring that up to the order $p+q$, the coefficients in
the Taylor expansion of the ratio of polynomials are equal to the
coefficients of the given power series. In the present contribution, we
take as this power series the perturbation expansion (\ref{eqC3}) in
the parameter $\beta$ that is put equal to one in the final formulas.
We note that a different power series that is not explicitly defined,
seems to have been used for the Pad\'e approximants in
\cite{KucharskiNogaBartlett89}. For the application of rational
approximants to the M{\o}ller-Plesset series  see also Ref.\
\cite{HandyKnowlesSomasundram85}.

                     \section{Numerical Results}
Fortunately, excellent data for the test of the methods
described in the previous section are available in
\cite{SchmidtWarkenHandy93}. This paper also includes results given in
\cite{KucharskiNogaBartlett89}. In these references, a large
number of M{\o}ller-Plesset results up to fifth order, and FCI
(Full Configuration Interaction) or CCSDT (Coupled Cluster
Singles Doubles Triples) results are given for the ground states
of benchmark molecules (BH, HF, CH${}_2$, H${}_2$O, NH${}_2$,
NH${}_3$, CO, C${}_2$H${}_2$, O${}_3$, CN). The results of the
reanalysis of these data is presented in Table \ref{tab1}. For
completeness, the MP data are also plotted. If not stated
otherwise, MPn means RMPn in open shell cases. Apart from case
n (NH${}_3$), the left half of the data in Table \ref{tab1} is obtained
from the data up to fourth order, while the right half also
depends on the fifth order.

It is seen that in many cases, the correlation energy estimators
provide excellent results. Problematic cases are s, t, and u. In case
s corresponding to CN, the perturbation series is divergent, being based on doubly occupied
ROHF orbitals where for alpha and beta spins the same orbitals are
used, unlike the RMP orbitals where occupied alpha and beta set both
are rotated. \cite{SchmidtWarkenHandy93,Handy94} In cases t and
u corresponding to H${}_2$O at stretched geometries, the
approach is based on an UMP series that is monotonously and very slowly
convergent \cite{SchmidtWarkenHandy93,HandyKnowlesSomasundram85}.

\setlongtables
\setlength{\LTleft}{0pt}
\setlength{\LTright}{0pt}
\begin{longtable}{l@{\extracolsep{\fill}}rr|lrr}
\caption{Comparison of Correlation Energy Estimators}\label{tab1}    \\
 Method & \multicolumn{1}{c}{Energy} & \%Corr
& Method & \multicolumn{1}{c}{Energy} &
\%Corr \\
\hline
\endfirsthead
\multicolumn{6}{l}{(Table \ref{tab1} -- continued)}
\endhead
\multicolumn{6}{c}{Case a: BH (${}^1\Sigma$, $r=2.329\,a_0$, DZP,
\cite{KucharskiNogaBartlett89,HarrisonHandy83,%
BartlettSekinoPurvis83})} \\
 SCF    &  -25.125260 &   0.00   & MP5    &  -25.225101 &  97.53 \\
 MP2    &  -25.198988 &  72.02   & F5     &  -25.226881 &  99.27 \\
 MP3    &  -25.216566 &  89.19   & GF5    &  -25.226971 &  99.36 \\
 MP4    &  -25.222567 &  95.06   & GF5b   &  -25.227088 &  99.47 \\
 F4     &  -25.226167 &  98.57   & [3,2]  &  -25.227299 &  99.68 \\
 $[2,2]$&  -25.225294 &  97.72   & [2,3]  &  -25.227478 &  99.85 \\
 $\Pi$2 &  -25.226555 &  98.95   & FCI    &  -25.227627 & 100.00 \\
\hline
\multicolumn{6}{c}{Case b: BH (${}^1\Sigma$, $r=1.5 \times 2.329\,a_0$, DZP,
\cite{KucharskiNogaBartlett89,NogaBartlett87})} \\
 SCF    &  -25.062213 &   0.00   & MP5    &  -25.172372 &  96.83 \\
 MP2    &  -25.139869 &  68.26   & F5     &  -25.174484 &  98.69 \\
 MP3    &  -25.160249 &  86.18   & GF5    &  -25.174544 &  98.74 \\
 MP4    &  -25.168745 &  93.64   & GF5b   &  -25.177010 & 100.91 \\
 F4     &  -25.175345 &  99.45   & [3,2]  &  -25.175078 &  99.21 \\
 $[2,2]$&  -25.173623 &  97.93   & [2,3]  &  -25.175106 &  99.24 \\
 $\Pi$2 &  -25.176791 & 100.72   & FCI    &  -25.175976 & 100.00 \\
\hline
\multicolumn{6}{c}{Case c: BH (${}^1\Sigma$, $r=2 \times 2.329\,a_0$, DZP,
\cite{KucharskiNogaBartlett89,NogaBartlett87})} \\
 SCF    &  -24.988201 &   0.00   & MP5    &  -25.121278 &  95.65 \\
 MP2    &  -25.074503 &  62.03   & F5     &  -25.126844 &  99.65 \\
 MP3    &  -25.100221 &  80.51   & GF5    &  -25.126983 &  99.75 \\
 MP4    &  -25.114005 &  90.42   & GF5b   &  -25.130104 & 101.99 \\
 F4     &  -25.128829 & 101.08   & [3,2]  &  -25.129407 & 101.49 \\
 $[2,2]$&  -25.124953 &  98.29   & [2,3]  &  -25.129475 & 101.54 \\
 $\Pi$2 &  -25.137084 & 107.01   & FCI    &  -25.127333 & 100.00 \\
\hline
\multicolumn{6}{c}{Case d: HF ($r=1.733 \,a_0$, DZP,
\cite{KucharskiNogaBartlett89,BauschlicherLanghoffTaylorHandyKnowles86})} \\
 SCF    & -100.047087 &   0.00   & MP5    & -100.250158 &  99.60 \\
 MP2    & -100.243165 &  96.17   & F5     & -100.250099 &  99.57 \\
 MP3    & -100.245531 &  97.33   & GF5    & -100.250276 &  99.66 \\
 MP4    & -100.251232 & 100.13   & GF5b   & -100.251988 & 100.50 \\
 F4     & -100.251443 & 100.23   & [3,2]  & -100.250468 &  99.75 \\
 $[2,2]$& -100.251547 & 100.28   & [2,3]  & -100.250481 &  99.76 \\
 $\Pi$2 & -100.251820 & 100.42   & FCI    & -100.250969 & 100.00 \\
\hline
\multicolumn{6}{c}{Case e: HF ($r=1.5 \times 1.733 \,a_0$, DZP,
\cite{KucharskiNogaBartlett89,BauschlicherLanghoffTaylorHandyKnowles86})} \\
 SCF    &  -99.933230 &   0.00   & MP5    & -100.158121 &  99.00 \\
 MP2    & -100.149756 &  95.32   & F5     & -100.158152 &  99.01 \\
 MP3    & -100.148543 &  94.78   & GF5    & -100.158247 &  99.05 \\
 MP4    & -100.159627 &  99.66   & GF5b   & -100.161609 & 100.53 \\
 F4     & -100.159443 &  99.58   & [3,2]  & -100.158750 &  99.28 \\
 $[2,2]$& -100.160091 &  99.87   & [2,3]  & -100.158757 &  99.28 \\
 $\Pi$2 & -100.160708 & 100.14   & FCI    & -100.160395 & 100.00 \\
\hline
\multicolumn{6}{c}{Case f: HF ($r=2 \times 1.733 \,a_0$, DZP,
\cite{KucharskiNogaBartlett89,BauschlicherLanghoffTaylorHandyKnowles86})} \\
 SCF    &  -99.817571 &   0.00   & MP5    & -100.073004 &  96.93 \\
 MP2    & -100.057062 &  90.88   & F5     & -100.073139 &  96.98 \\
 MP3    & -100.054148 &  89.77   & GF5    & -100.073301 &  97.04 \\
 MP4    & -100.076267 &  98.16   & GF5b   & -100.079678 &  99.46 \\
 F4     & -100.075480 &  97.86   & [3,2]  & -100.075064 &  97.71 \\
 $[2,2]$& -100.077899 &  98.78   & [2,3]  & -100.075072 &  97.71 \\
 $\Pi$2 & -100.080476 &  99.76   & FCI    & -100.081107 & 100.00 \\
\hline
\multicolumn{6}{c}{Case g: CH${}_2$ (${}^1A_1$, $r=2.11 \,a_0$,
$\theta=102.4\,{}^{\circ}$, DZP,
\cite{KucharskiNogaBartlett89,BauschlicherTaylor86b})} \\
 SCF    &  -38.886297 &   0.00   & MP5    &  -39.024234 &  97.91 \\
 MP2    &  -38.996127 &  77.96   & F5     &  -39.025336 &  98.69 \\
 MP3    &  -39.016593 &  92.48   & GF5    &  -39.025450 &  98.77 \\
 MP4    &  -39.022203 &  96.47   & GF5b   &  -39.025413 &  98.74 \\
 F4     &  -39.024615 &  98.18   & [3,2]  &  -39.025674 &  98.93 \\
 $[2,2]$&  -39.024049 &  97.78   & [2,3]  &  -39.025895 &  99.09 \\
 $\Pi$2 &  -39.024791 &  98.30   & FCI    &  -39.027183 & 100.00 \\
\hline
\multicolumn{6}{c}{Case h: H${}_2$O (${}^1A_1$, $r=1.88973 \,a_0$,
$\theta=104.5\,{}^{\circ}$, DZP,
\cite{KucharskiNogaBartlett89,BauschlicherTaylor86a})} \\
 SCF    &  -76.040542 &   0.00   & MP5    &  -76.255924 &  99.68 \\
 MP2    &  -76.243660 &  94.00   & F5     &  -76.255918 &  99.67 \\
 MP3    &  -76.249403 &  96.66   & GF5    &  -76.255929 &  99.68 \\
 MP4    &  -76.255706 &  99.58   & GF5b   &  -76.257338 & 100.33 \\
 F4     &  -76.256262 &  99.83   & [3,2]  &  -76.256134 &  99.77 \\
 $[2,2]$&  -76.256282 &  99.84   & [2,3]  &  -76.256135 &  99.77 \\
 $\Pi$2 &  -76.256729 & 100.05   & FCI    &  -76.256624 & 100.00 \\
\hline
\multicolumn{6}{c}{Case i: H${}_2$O (${}^1A_1$, $r=1.5\times 1.88973 \,a_0$,
$\theta=104.5\,{}^{\circ}$, DZP,
\cite{KucharskiNogaBartlett89,BauschlicherTaylor86a})} \\
 SCF    &  -75.800494 &   0.00   & MP5    &  -76.066422 &  98.16 \\
 MP2    &  -76.048095 &  91.40   & F5     &  -76.066368 &  98.14 \\
 MP3    &  -76.045081 &  90.28   & GF5    &  -76.066442 &  98.17 \\
 MP4    &  -76.065641 &  97.87   & GF5b   &  -76.068395 &  98.89 \\
 F4     &  -76.064909 &  97.60   & [3,2]  &  -76.068528 &  98.94 \\
 $[2,2]$&  -76.066937 &  98.35   & [2,3]  &  -76.068533 &  98.94 \\
 $\Pi$2 &  -76.068954 &  99.10   & FCI    &  -76.071405 & 100.00 \\
\hline
\multicolumn{6}{c}{Case j: H${}_2$O (${}^1A_1$, $r=2\times 1.88973 \,a_0$,
$\theta=104.5\,{}^{\circ}$, DZP,
\cite{KucharskiNogaBartlett89,BauschlicherTaylor86a})} \\
 SCF    &  -75.582286 &   0.00   & MP5    &  -75.935304 &  95.41 \\
 MP2    &  -75.898603 &  85.50   & F5     &  -75.934525 &  95.20 \\
 MP3    &  -75.877664 &  79.84   & GF5    &  -75.935353 &  95.43 \\
 MP4    &  -75.937410 &  95.98   & GF5b   &  -75.923566 &  92.24 \\
 F4     &  -75.927115 &  93.20   & [3,2]  &  -75.949379 &  99.22 \\
 $[2,2]$&  -75.941045 &  96.97   & [2,3]  &  -75.949401 &  99.22 \\
 $\Pi$2 &  -75.954930 & 100.72   & FCI    &  -75.952269 & 100.00 \\
\hline
\multicolumn{6}{c}{Case k: NH${}_2$ (${}^2B_1$, $r=1.013 \,\mbox{\AA}$,
$\theta=103.2\,{}^{\circ}$, 6-31G,
\cite{HandyKnowlesSomasundram85,KnowlesAndrewsAmosHandyPople91})} \\
 SCF    &  -55.530177 &   0.00   & MP5    &  -55.632426 &  99.18 \\
 MP2    &  -55.617272 &  84.48   & F5     &  -55.632818 &  99.56 \\
 MP3    &  -55.627501 &  94.40   & GF5    &  -55.632834 &  99.57 \\
 MP4    &  -55.631220 &  98.01   & GF5b   &  -55.633280 & 100.00 \\
 F4     &  -55.632525 &  99.27   & [3,2]  &  -55.633011 &  99.74 \\
 $[2,2]$&  -55.632204 &  98.96   & [2,3]  &  -55.633022 &  99.75 \\
 $\Pi$2 &  -55.632825 &  99.56   & FCI    &  -55.633276 & 100.00 \\
\hline
\multicolumn{6}{c}{Case l: NH${}_2$ (${}^2B_1$, $r=1.5\times 1.013 \,\mbox{\AA}$,
$\theta=103.2\,{}^{\circ}$, 6-31G,
\cite{HandyKnowlesSomasundram85,KnowlesAndrewsAmosHandyPople91})} \\
 SCF    &  -55.367729 &   0.00   & MP5    &  -55.520522 &  96.14 \\
 MP2    &  -55.489967 &  76.91   & F5     &  -55.521721 &  96.89 \\
 MP3    &  -55.504270 &  85.91   & GF5    &  -55.521724 &  96.90 \\
 MP4    &  -55.516470 &  93.59   & GF5b   &  -55.523319 &  97.90 \\
 F4     &  -55.521456 &  96.73   & [3,2]  &  -55.523696 &  98.14 \\
 $[2,2]$&  -55.521125 &  96.52   & [2,3]  &  -55.523706 &  98.14 \\
 $\Pi$2 &  -55.526202 &  99.71   & FCI    &  -55.526658 & 100.00 \\
\hline
\multicolumn{6}{c}{Case m: NH${}_2$ (${}^2B_1$, $r=2\times 1.013 \,\mbox{\AA}$,
$\theta=103.2\,{}^{\circ}$, 6-31G,
\cite{HandyKnowlesSomasundram85,KnowlesAndrewsAmosHandyPople91})} \\
 SCF    &  -55.181593 &   0.00   & MP5    &  -55.418215 &  91.36 \\
 MP2    &  -55.357617 &  67.96   & F5     &  -55.420149 &  92.11 \\
 MP3    &  -55.375463 &  74.85   & GF5    &  -55.420173 &  92.12 \\
 MP4    &  -55.409165 &  87.87   & GF5b   &  -55.412429 &  89.13 \\
 F4     &  -55.421427 &  92.60   & [3,2]  &  -55.432093 &  96.72 \\
 $[2,2]$&  -55.426946 &  94.73   & [2,3]  &  -55.432101 &  96.72 \\
 $\Pi$2 &  -55.478348 & 114.58   & FCI    &  -55.440593 & 100.00 \\
\hline
\multicolumn{6}{c}{Case n: NH${}_3$ ($r=1.91165 \,a_0$,
$\theta=106.7\,{}^{\circ}$, DZ,
\cite{HarrisonHandy83,BartlettSekinoPurvis83})} \\
 SCF    &  -56.165931 &   0.00   & F4     &  -56.291937 &  99.47 \\
 MP2    &  -56.277352 &  87.95   & $[2,2]$&  -56.291782 &  99.35 \\
 MP3    &  -56.285281 &  94.21   & $\Pi$2 &  -56.292636 & 100.02 \\
 MP4    &  -56.290692 &  98.48   & FCI    &  -56.292612 & 100.00 \\
\hline
\multicolumn{6}{c}{Case o: CO (${}^1\Sigma$,
DZ,
\cite{KucharskiNogaBartlett89})} \\
 SCF    & -112.760093 &   0.00   & MP5    & -113.059117 &  98.36 \\
 MP2    & -113.045824 &  93.99   & F5     & -113.059254 &  98.41 \\
 MP3    & -113.044659 &  93.61   & GF5    & -113.060859 &  98.93 \\
 MP4    & -113.067749 & 101.20   & GF5b   & -113.073579 & 103.12 \\
 F4     & -113.067469 & 101.11   & [3,2]  & -113.062479 &  99.47 \\
 $[2,2]$& -113.069566 & 101.80   & [2,3]  & -113.062539 &  99.49 \\
 $\Pi$2 & -113.072074 & 102.62   & CCSDT  & -113.064100 & 100.00 \\
\hline
\multicolumn{6}{c}{Case p: C${}_2$H${}_2$ (${}^1\Sigma_g$,
DZP,
\cite{KucharskiNogaBartlett89})} \\
 SCF    &  -76.831819 &   0.00   & MP5    &  -77.118892 & 102.18 \\
 MP2    &  -77.085307 &  90.23   & F5     &  -77.120192 & 102.65 \\
 MP3    &  -77.097232 &  94.47   & GF5    &  -77.122141 & 103.34 \\
 MP4    &  -77.111732 &  99.63   & GF5b   &  -77.117205 & 101.58 \\
 F4     &  -77.113928 & 100.42   & [3,2]  &  -77.127079 & 105.10 \\
 $[2,2]$&  -77.114110 & 100.48   & [2,3]  &  -77.127731 & 105.33 \\
 $\Pi$2 &  -77.116235 & 101.24   & CCSDT  &  -77.112760 & 100.00 \\
\hline
\multicolumn{6}{c}{Case q: O${}_3$ (${}^1A_1$,
DZP,
\cite{KucharskiNogaBartlett89})} \\
 SCF    & -224.295920 &   0.00   & MP5    & -224.929902 &  97.54 \\
 MP2    & -224.931924 &  97.86   & F5     & -224.933812 &  98.15 \\
 MP3    & -224.888104 &  91.11   & GF5    & -224.934513 &  98.25 \\
 MP4    & -224.952784 & 101.07   & GF5b   & -224.952167 & 100.97 \\
 F4     & -224.941418 &  99.32   & [3,2]  & -224.938301 &  98.84 \\
 $[2,2]$& -224.950280 & 100.68   & [2,3]  & -224.938367 &  98.85 \\
 $\Pi$2 & -224.952387 & 101.00   & CCSDT  & -224.945859 & 100.00 \\
\hline
\multicolumn{6}{c}{Case r: CN (${}^2\Sigma$, $r=1.1619\, \,\mbox{\AA}$,
STO-3G, RMP
\cite{KnowlesAndrewsAmosHandyPople91})} \\
 SCF    &  -90.99752 &   0.00    & MP5    &  -91.16157 &  95.07 \\
 MP2    &  -91.15437 &  90.90    & F5     &  -91.16165 &  95.12 \\
 MP3    &  -91.14799 &  87.20    & GF5    &  -91.16166 &  95.12 \\
 MP4    &  -91.16300 &  95.90    & GF5b   &  -91.16360 &  96.24 \\
 F4     &  -91.16133 &  94.93    & [3,2]  &  -91.16297 &  95.88 \\
 $[2,2]$&  -91.16321 &  96.02    & [2,3]  &  -91.16297 &  95.88 \\
 $\Pi$2 &  -91.16426 &  96.63    & FCI    &  -91.17008 & 100.00 \\
\hline
\multicolumn{6}{c}{Case s: CN (${}^2\Sigma$, $r=1.1619\, \,\mbox{\AA}$,
STO-3G, Hubac-Carsky,
\cite{Handy94,HubacCarsky80})} \\
 SCF    &  -90.99752  &   0.00   & MP5    &  -91.12039  &  71.20 \\
 MP2    &  -91.17762  & 104.37   & F5     &  -91.15212  &  89.59 \\
 MP3    &  -91.14160  &  83.50   & GF5    &  -91.15998  &  94.15 \\
 MP4    &  -91.19422  & 113.99   & GF5b   &  -91.18190  & 106.85 \\
 F4     &  -91.17389  & 102.21   & [3,2]  &  -91.16350  &  96.19 \\
 $[2,2]$&  -91.18753  & 110.11   & [2,3]  &  -91.16359  &  96.24 \\
 $\Pi$2 &  -91.19152  & 112.42   & FCI    &  -91.17008  & 100.00 \\
\hline
\multicolumn{6}{c}{Case t: H${}_2$O  ($r=1.5 \times 0.967\, \,\mbox{\AA}$,
$\theta=107.6\,{}^{\circ}$,
6-21G,\cite{HandyKnowlesSomasundram85})} \\
 RHF    &  -75.707206 &   0.00        & UMP5    &  -75.853895 & 76.41 \\
 UHF    &  -75.735012 &  14.48        & F5      &  -75.855560 & 77.28 \\
 UMP2   &  -75.829388 &  63.65        & GF5     &  -75.856608 & 77.82 \\
 UMP3   &  -75.836823 &  67.52        & GF5b    &  -75.850870 & 74.84 \\
 UMP4   &  -75.848211 &  73.45        & [3,2]   &  -75.862349 & 80.81 \\
 F4     &  -75.851276 &  75.05        & [2,3]   &  -75.862421 & 80.85 \\
 $[2,2]$&  -75.851994 &  75.42        & FCI     &  -75.899180 &100.00 \\
 $\Pi$2 &  -75.857074 &  78.07        &         &             & \\
\hline
\multicolumn{6}{c}{Case u: H${}_2$O  ($r=2 \times 0.967\, \,\mbox{\AA}$,
$\theta=107.6\,{}^{\circ}$,
6-21G,\cite{HandyKnowlesSomasundram85})} \\
 RHF    &  -75.491406 &  0.00         & UMP5    &  -75.763370 & 90.72 \\
 UHF    &  -75.699298 & 69.35         & F5      &  -75.763704 & 90.83 \\
 UMP2   &  -75.754669 & 87.82         & GF5     &  -75.763826 & 90.88 \\
 UMP3   &  -75.760219 & 89.67         & GF5b    &  -75.763657 & 90.82 \\
 UMP4   &  -75.762422 & 90.41         & [3,2]   &  -75.764089 & 90.96 \\
 F4     &  -75.763098 & 90.63         & [2,3]   &  -75.764104 & 90.97 \\
 $[2,2]$&  -75.762941 & 90.58         & FCI     &  -75.791180 &100.00 \\
 $\Pi$2 &  -75.763281 & 90.69         &         &             & \\
\hline
\hline
\end{longtable}

Apart
from these problematic cases, it is seen that in case
m corresponding to NH${}_2$ at twice the equilibrium distances, the errors are
rather high. Excluding this case also, one may study the performance of
the correlation energy estimators statistically as shown in Table
\ref{tab1a}. Plotted are the maximal error, the mean absolute error,
the root mean square (rms) absolute error, and the mean percentage of
the correlation energy as obtained with the various methods. In
cases o, p, and q corresponding to the molecules CO,
C${}_2$H${}_2$, O${}_3$, respectively, no FCI result is
available. The statistical comparison is done
once excluding these cases, and once including these cases where as
reference for the error calculation the CCSDT result is taken.
For these cases, the given correlation energies
should thus be taken with care. Carefully designed fourth order methods like
$\Pi$2 yield correlation energy estimates that can compete with fifth
order results. As regards the fifth order methods, it seems that the
Goldhammer-Feenberg estimator GF5 is slightly superior to the Feenberg
energy F5, and the somewhat {\em ad hoc} estimator GF5b performs
surprisingly well. Among the Pad\'e approximants, the $[3,2]$ approximant (that
is invariant under the Feenberg scaling) is a rather successful
correlation estimator while the $[2,3]$ approximant performs very
similarly. Other Pad\'e approximants (not displayed in Table
\ref{tab1}) do not perform as well as the ones given in this table when
applied to the same data.

\begin{table}
\caption{Statistical comparison of various correlation energy
estimators}\label{tab1a}
\begin{tabular*}{\linewidth}{@{}l@{\extracolsep{\fill}}rrrr@{}}
\hline
Method   & max $\vert error\vert$   & mean $\vert error\vert$   & rms
$\vert error\vert$   & mean $\%$Corr\\
\hline
\multicolumn{5}{@{}c@{}}{Sampling 14 cases (a-l,n,r)}\\
F4      & 0.02515   & 0.00433 & 0.00767 &  98.3\\
$[2,2]$ & 0.01122   & 0.00319 & 0.00433 &  98.3\\
$\Pi$2  & 0.00975   & 0.00199 & 0.00329 & 100.1\\
\hline
\multicolumn{5}{@{}c@{}}{Sampling 17 cases (a-l,n-r)}\\
F4      & 0.02515   & 0.00409 & 0.00710 &  98.6\\
$[2,2]$ & 0.01122   & 0.00329 & 0.00430 &  98.8\\
$\Pi$2  & 0.00975   & 0.00269 & 0.00398 & 100.3\\
\hline
\multicolumn{5}{@{}c@{}}{Sampling 13 cases (a-l,r)}\\
F5      & 0.01774   & 0.00407 & 0.00628 &  98.2\\
GF5     & 0.01692   & 0.00394 & 0.00607 &  98.2\\
GF5b    & 0.02870   & 0.00400 & 0.00834 &  99.0\\
$[3,2]$ & 0.00711   & 0.00228 & 0.00308 &  99.1\\
$[2,3]$ & 0.00711   & 0.00224 & 0.00307 &  99.1\\
\hline
\multicolumn{5}{@{}c@{}}{Sampling 16 cases (a-l,o-r)}\\
 F5      & 0.01774 & 0.00483 & 0.00678 & 98.5\\
 GF5     & 0.01692 & 0.00470 & 0.00664 & 98.6\\
 GF5b    & 0.02870 & 0.00452 & 0.00811 & 99.6\\
 $[3,2]$ & 0.01432 & 0.00332 & 0.00492 & 99.4\\
 $[2,3]$ & 0.01497 & 0.00332 & 0.00503 & 99.5\\
\hline
\end{tabular*}
\end{table}

A careful analysis of the data in Table \ref{tab1} reveals that the
correlation energy estimation based on MP perturbation theory is the
better the closer one is to the optimal geometries of the molecule
under consideration. This is not very much surprising since it is
well-known that the quality of the MP series deteriorates with
increasing separations from the equilibrium geometries. Compare for
instance the triples of cases (a,b,c) for BH, (d,e,f) for HF,
(h,i,j) for H${}_2$O, and (k,l,m) for NH${}_2$,
with ratios 1:1.5:2 of the relevant distances. The values away from the
equilibrium geometries may or may not be reliable. The data, however,
suggest that then the correlation energy estimates are reliable if ---
as in cases f for HF at $2\times r_e$ and i for H${}_2$O at
$1.5\times r_e$--- the values of $\Pi$2, F4 and $[2,2]$ do not
differ too much from each other. In this situation, the $\Pi$2
estimator seems to provide the best results. On the other hand, large
differences between the estimates $\Pi$2, F4 and $[2,2]$ --- as in the
cases j for H${}_2$O at $2\times r_e$ and m for NH${}_2$ at
$2\times r_e$ --- clearly indicate that in these cases more
sophisticated methods (for instance the $\Lambda$ transformation
\cite{SchmidtWarkenHandy93,DietzSchmidtWarkenHess92,%
DietzSchmidtWarkenHess93a,DietzSchmidtWarkenHess93b,%
DietzSchmidtWarkenHess93c,DietzSchmidtWarkenHess94a,%
DietzSchmidtWarkenHess94b} or multi-reference methods
\cite{JeziorskiMonkhorst81,Hose89,SteinbornSilver96,%
Lindgren74,WahlDas77,%
Werner87,MurphyMessmer93}) are needed to
calculate the correlation energies reliably. As regards the fifth order
estimates, it is similarly seen that a large spread of the values of
the various estimates reveals that the MP based methods do not provide
sufficiently accurate results. Reversely, a small spread of the various
estimates indicates that with a high probability, the (R)MP based
correlation energy estimates are reliable.

Comparing fourth and fifth order based estimators, it is seen that the
latter do not always provide better estimates of the correlation
energy. In many cases, the $\Pi$2 estimate that is based on fourth
order, provides results of comparable quality.

\begin{table}
\caption{Dissociation barrier (kJ/mol) of H${}_2$CO$\longrightarrow$H${}_2+{}$CO
using a TZ2P basis at MP2 geometries ${}^{a}$}\label{tab2}
\begin{tabular*}{\linewidth}{@{}l@{\extracolsep{\fill}}...l@{}}
\hline
 Method &  \multicolumn{1}{c}{Minimum}     &
 \multicolumn{1}{c}{Transition state} & \multicolumn{1}{c@{}}{Barrier
} & Ref. \\
\hline
 SCF     &  -113.912879 & -113.748693 &  431.1 & \cite{SchmidtWarkenHandy93}\\
 MP2     &  -114.329202 & -114.182435 &  385.3 & \cite{SchmidtWarkenHandy93}\\
 MP3     &  -114.334186 & -114.185375 &  390.7 & \cite{SchmidtWarkenHandy93}\\
 MP4     &  -114.359894 & -114.219892 &  367.6 & \cite{SchmidtWarkenHandy93}\\
 F4      &  -114.360838 & -114.220603 &  368.2 & \cite{SchmidtWarkenHandy93}\\
 $[2,2]$ &  -114.362267 & -114.223409 &  364.6 & This work\\
 $\Pi$2  &  -114.364840 & -114.227767 &  359.9 & This work\\
 BE${}^{b}$
         &              &             &  360   & \cite{DupuisLesterLengsfieldLiu83}\\
\hline
\end{tabular*}
\\
${}^{a}$ \cite{SchmidtWarkenHandy93} \\
${}^{b}$ Best estimate \cite{DupuisLesterLengsfieldLiu83}
\end{table}

\begin{table}%
\caption{Barrier height and heat of reaction (kJ/mol) for
CH${}_3+{}$C${}_2$H${}_4\longrightarrow{}$C${}_3$H${}_7$ with a
6-31G${}^{*}$
basis${}^{a}$}\label{tab3}%
\begin{tabular*}{\linewidth}{@{}l@{\extracolsep{\fill}}rrrrl@{}}
\hline
Method    &  \multicolumn{1}{l}{Reactants}      &
\multicolumn{1}{l}{TS${}^{b}$} & \multicolumn{1}{l}{Product}         &
\multicolumn{1}{c}{Barrier} & \multicolumn{1}{c@{}}{HR${}^{c}$}\\
\hline
 RHF      &  $-$117.585674  &  $-$117.553736  &  $-$117.626572  & 83.8 & $-$107.4 \\
 RMP2     &  $-$117.967150  &  $-$117.952092  &  $-$118.014126  & 39.5 & $-$123.3 \\
 RMP3     &  $-$118.004259  &  $-$117.986543  &  $-$118.049999  & 46.5 & $-$120.1 \\
 RMP4     &  $-$118.022888  &  $-$118.008072  &  $-$118.066816  & 38.9 & $-$115.3 \\
 F4       &  $-$118.028674  &  $-$118.014137  &  $-$118.071720  & 38.2 & $-$113.0 \\
 $[2,2]$  &  $-$118.027529  &  $-$118.013226  &  $-$118.070703  & 37.6 & $-$113.3 \\
 $\Pi$2   &  $-$118.030923  &  $-$118.017302  &  $-$118.073432  & 35.8 & $-$111.6 \\
 exp.${}^{d}$     &                 &                 &                 & 33.1 & $-$107   \\
\hline
\end{tabular*}
\\
${}^{a}$ \cite{SchmidtWarkenHandy93} \\
${}^{b}$ Transition state\\
${}^{c}$ Heat of reaction\\
${}^{d}$ \cite{Kerr72,XXX85,CastelhanoGriller82,SchmidtWarkenHandy93}
\end{table}

In Tables \ref{tab2} and \ref{tab3} the correlation energy
estimators are used to calculate the dissociation barrier for
H${}_2$CO$\longrightarrow$H${}_2+{}$CO, and the barrier height
and the heat of reaction for
CH${}_3+{}$C${}_2$H${}_4\longrightarrow{}$C${}_3$H${}_7$.

In both examples, the calculation is based on known M{\o}ller-Plesset
energies up to fourth order \cite[Tab. 2-4]{SchmidtWarkenHandy93}. The
results show that reliable correlation energy estimates as provided by
the Feenberg energy F4 \cite{SchmidtWarkenHandy93}, the Pad\'e
approximant $[2,2]$, and the effective characteristic polynomial
estimate $\Pi$2 lead to good agreement with experimental data. The
$\Pi$2 estimator yields in both cases the best results.

In summary, it has been shown that the availability of various
estimators based on (R)MP results allows in many cases the
accurate calculation of the correlation energy at negligible
additional computational costs. Also, larger deviations between
the values indicate clearly cases where further work is
necessary.

Finally, we note that the above estimators are expected to be
useful to improve convergence of perturbation series for the
energies also for the multi-reference case. This conjecture is
a promising topic for further investigations.

\begin{ack}
I thank Dr. E. J. Weniger and Prof. Dr. J. {{\v C}{\' \i}{\v
z}ek} for discussions regarding the effective characteristic
polynomial approach. I am grateful to Prof.\ Dr.\
E.\ O.\ Steinborn for his support and the excellent working
conditions at Regensburg. Help of the staff of the computing
centre of the University  of Regensburg is thankfully
acknowledged.
\end{ack}

\end{document}